\documentclass[aps,prd,superscriptaddress,showpacs,preprint,amsmath,amssymb]{revtex4}
\usepackage{graphicx, bm}

\usepackage[usenames]{color}
\usepackage{slashed}

\usepackage{subcaption}
\usepackage{slashed}
\usepackage{float}
\begin{document}

\draft

\title{Probing the anomalous triple $ZZ\gamma$ and $Z\gamma\gamma $ couplings at the FCC-$\mu p$ and SPPC-$\mu p$}

\author{Emre Gurkanli\footnote{egurkanli@sinop.edu.tr}}
\affiliation{\small Department of Physics, Sinop University, Turkey.\\}

\date{\today}

\begin{abstract}

In this study, 24.5 TeV CoM energy FCC-$\mu p$ and 20.2 TeV CoM energy SPPC-$\mu p$ muon-proton colliders have been utilized to explore the anomalous $ZZ\gamma$ and $Z\gamma\gamma$ couplings corresponding to dim-8 operators through the process of $\mu^- \gamma \to Z l^{-} \to l^{-} \tilde{\nu_{l}} \nu_{l}$. A cut-based method has been applied to enhance the signal background ratio in the analysis. Coupling sensitivities, at a 95\% Confidence Level (C.L.), under the systematic uncertainties of 0\% and 5\%, were obtained with luminosities of ${\cal L}_{int}=5$ and 42.8 fb$^{-1}$ for FCC-$\mu p$ and SPPC-$\mu p$ colliders, respectively. The sensitivities for anomalous $C_{BB}/\Lambda^{4}$, $C_{\tilde{B}W}/\Lambda^{4}$, $C_{BW}/\Lambda^{4}$ and $C_{WW}/\Lambda^{4}$ couplings without systematic uncertainty for FCC-$\mu p$ and SPPC-$\mu p$ were found as follows, respectively: [0.068; 0.070] TeV-4, [-0.187; 0.184] TeV-4, [-0.236; 0.238] TeV-4, [-0.619; 0.621] TeV-4, and [-0.066; 0.051] TeV-4, [-0.158; 0.151] TeV-4, [-0.207; 0.192] TeV-4, [-0.514; 0.522] TeV-4.

\end{abstract}


\vspace{5mm}

\maketitle


\section{Introduction}

The Standard Model is a successful theory that interprets fundamental particles' behavior and interactions within reachable energy constraints in present collider experiments. However, a more comprehensive theory is required to address unresolved issues in the universe, such as the strong CP problem, non-zero neutrino masses, and the baryon asymmetry in the early universe. The non-Abelian gauge symmetry of the SM describes the self-interactions of the gauge bosons. These interactions can be defined by triple gauge boson couplings, $WWV$,$ZZV$, and $ZV\gamma$ (V=$\gamma, Z$) \cite{Baur:2000hfg}. However, interactions involving the photon and $Z$-boson are not present at the lowest order in the SM, as the $Z$-boson has no electric charge. The absence of $ZZ\gamma$, $Z\gamma\gamma$, and $ZZZ$ triple interactions in the SM leads to deviations from SM predictions in the presence of these vertices, providing sensitive evidence for new physics. In the literature, aTGC have been extensively investigated in various production processes in $ee$ \cite{Spor:2020wft,Billur:2019cav,Ari:2015tca,Choudhury:1994ywq,Atag:2004ybz,Ots:2004twm,Ots:2006gsd,Rodriguez:2009rnw,Ananthanarayan:2012onz,Ananthanarayan:2014cal,Rahaman:2016nzs,Rahaman:2017qed,Ellis:2020ekm,Fu:2021jec,Ellis:2021rop,Yang:2022tgw,Spor:2022pou,Jahedi:2022duc,Jahedi:2023myu} and $pp$ \cite{Baur:1993fkx,Senol:2018gvg,Rahaman:2019tnp,Senol:2019ybv,Senol:2020hbh,Yilmaz:2020ser,Yilmaz:2021dbm,Hernandez:2021wsz,Biekotter:2021ysx,Lombardi:2022plb,Hernandez-Juarez:2022kjx,Ellis:2022zdw,Ellis:2023ucy,Chapon:2009hh,Kepka:2008yx} colliders. Studies in accelerator physics that contain a variety of collision types play a crucial role in advancing research on new physics within particle physics \cite{Geng:2019ebo,Gounaris:2003lsd,Belloni:2022due,Acar:2016rde, Canbay:2017rbg, Ketenoglu:2022fzo}. The LHC is a high-potential Hadron collider that explores new particles and interactions. However, the final state jets scattered after the collision of proton beams lead to complex backgrounds, making it challenging to make precise measurements at the LHC to detect the signals. As the most potent and expansive circular proton-proton collider ever constructed, the LHC is set to undergo gradual enhancements through advancing accelerator technology. Pursuing novel physics beyond the Standard Model (SM), lepton-hadron colliders are promising contenders in the trajectory of particle physics. After the LHC era, the strategy entails converting the LHC into the LHeC by adding an electron ring adjacent to the primary LHC tunnel. Subsequent phases involve substituting the electron ring with a muon counterpart, establishing a new lepton-hadron collider named LHC-$\mu p$. On the other hand, the FCC is planned to be a circular collider at CERN for the post-LHC epoch. The FCC initiative involves the conceptualization of a forthcoming $ee$ collider, with plans to incorporate $pp$, $ep$, $\mu\mu$, and $\mu p$ colliders. Notably, FCC-hh is envisioned as a future $pp$ collider boasting a CoM energy of 100 TeV \cite{Delahaye:2019egb}. Introducing a muon ring addition to the FCC will facilitate the formation of high energy $\mu p$ colliders \cite{Long:2021wja}. Simultaneously,  a $pp$ collider named the Super Proton Proton Collider (SPPC) with a CoM energy of 70 TeV was designed in China. Preceding the SPPC collider, the CEPC is planned as the initial phase utilizing the same tunnel and is envisaged as the future $ee$ collider. The CEPC/SPPC project includes $\mu\mu$, $ep$, and $\mu p$ collisions to be conducted in later years, similar to the FCC project \cite{CEPC}. Many phenomenological studies have been done in muon colliders \cite{Palmer:2014asd,Antonelli:2016ezx,Wang:2016rwe,Neuffer:2018zxp,Boscolo:2019ytr,Bogomilov:2020twm,Buttazzo:2018wmg,Koksal:2019lja,Costantini:2020tkp,Yin:2020gre,Ruhdorfer:2020tgx,Chiesa:2020yhn,Bandyopadhyay:2021lja,Han:2021hrq,Liu:2021gtr,Han:2021twq,Capdevilla:2021xku,Bottaro:2021res,Capdevilla:2021ooc,Huang:2021edc,Asadi:2021wsd,Han:2021pas,Franceschini:2021pol,Arakawa:2022mkr,Chiesa:2021tyr,Buttazzo:2021eka,Huang:2022vke,Spor:2022kyz,Yang:2022dbn,Forslund:2022unz}.

On the other hand, this study presents an investigation into the anomalous neutral triple gauge couplings (aNTGCs) involving the $ZZ\gamma$ and $Z\gamma\gamma$ vertices. This investigation is conducted particularly in the context of future high-energy muon-proton colliders, such as the Future Circular Collider (FCC-$\mu p$) and the Super Proton-Proton Collider (SPPC-$\mu p$). While current research has extensively examined aTGCs like $WW\gamma$, $ZZZ$, and $WWZ$, the $ZZ\gamma$ and $Z\gamma\gamma$ vertices have not been sufficiently explored with future collider technologies, and this study aims to address these gaps in the literature. By leveraging the advantages of muon-proton colliders, which combine high energy and clean initial states, this study contributes to the field by investigating these interactions with good precision. 
Additionally, there are some studies in the literature that utilize UV-completed models and matching to provide complementary insights\cite{Cepedello:2024ogz,Ellis:2024omd}. While our work primarily adopts a model-independent approach within the Effective Field Theory (EFT) framework, it builds upon the robust theoretical foundation established by such studies. Utilizing Monte Carlo simulations and the Effective Field Theory (EFT) framework, the study proposes new methodologies that could significantly enhance the sensitivity to these couplings, thereby advancing our expectations of anomalous gauge interactions in the context of future collider experiments. The FCC-$\mu p$ and SPPC-$\mu p$ colliders have high CoM energies and luminosity values. The values used for the FCC-$\mu p$ collider are $E_{\mu}=3$ TeV, $E_{p}=50$ TeV, ${\cal L}_{int}=5$ fb$^{-1}$, and for the SPPC-$\mu p$ collider, $E_{\mu}=1.5$ TeV, $E_{p}=68$ TeV, ${\cal L}_{int}=42.8$ fb$^{-1}$ \cite{Acar:2017eli,Aydin:2021iky,Caliskan:2018vep}.

This paper is structured as follows: Section II provides an overview of the dimension-8 operators in the effective field theory. Section III focuses on the process $\mu^- \gamma \to Z l^{-} \to l^{-} \tilde{\nu_{l}} \nu_{l}$ and discusses its relevance in the context of the ANTGC, aim at the event generation and analysis part of the process. Section IV focuses on the obtained sensitivities on aNTGC. Finally, Section V presents our conclusions.

\section{Effective Theory Approach}

The Effective Field Theory (EFT) is useful for investigating potential new physics phenomena beyond the SM. Exploring aNTGC within the framework of the SM gauge group necessitates the addition of higher-order dimension operators into the SM Lagrangian. These operators introduce anomalous couplings, influencing the effective vertices\cite{Rahaman:2016nzs}. The focus of this study lies in examining dim-8 operators that characterize aNTGC. Thus, the effective Lagrangian, including SM interactions and new physics contributions, is given by \cite{Degrande:2014ydn}.

\begin{eqnarray}
\label{eq.1}
{\cal L}^{\text{NTGC}}={\cal L}_{\text{SM}}+\sum_{i}\frac{C_i}{\Lambda^{4}}({\cal O}_i+{\cal O}_i^\dagger)
\end{eqnarray}

{\raggedright Here, $\Lambda$ represents the scale of new physics, while the coefficients $C_i$ denote dimensionless parameters associated with the new physics. The dim-8 operators ${\cal O}_i$ are expressed in the following equations.

\begin{eqnarray}
\label{eq.2}
{\cal O}_{\widetilde{B}W}=iH^{\dagger} \widetilde{B}_{\sigma\lambda}W^{\sigma\mu} \{D_\mu,D^\lambda \}H,
\end{eqnarray}
\begin{eqnarray}
\label{eq.3}
{\cal O}_{BW}=iH^\dagger B_{\sigma\mu}W^{\sigma\rho} \{D_\rho,D^\mu \}H,
\end{eqnarray}
\begin{eqnarray}
\label{eq.4}
{\cal O}_{WW}=iH^\dagger W_{\sigma\mu}W^{\sigma\rho} \{D_\rho,D^\mu \}H,
\end{eqnarray}
\begin{eqnarray}
\label{eq.5}
{\cal O}_{BB}=iH^\dagger B_{\sigma\mu}B^{\sigma\nu} \{D_\nu,D^\mu \}H
\end{eqnarray}

{\raggedright where}

\begin{eqnarray}
\label{eq.6}
B_{\mu\nu}=\left(\partial_\mu B_\nu - \partial_\nu B_\mu\right),
\end{eqnarray}
\begin{eqnarray}
\label{eq.7}
W_{\mu\nu}=\sigma^i\left(\partial_\mu W_\nu^i - \partial_\nu W_\mu^i + g\epsilon_{ijk}W_\mu^j W_\nu^k\right),
\end{eqnarray}

{\raggedright with $\langle \sigma^j\sigma^i\rangle=\delta^{ji}/2$ and}

\begin{eqnarray}
\label{eq.8}
D_\mu \equiv \partial_\mu - i\frac{g^\prime}{2}B_\mu Y - ig_W W_\mu^i\sigma^i.
\end{eqnarray}

In these equations, $B_{\mu \nu}$ and $W_{\mu \nu}$ represent the field strength tensors, $H$ is the Higgs field, and $D_{\mu}$ stands for the covariant derivative. On the other hand, ${\cal O}_{\widetilde{B}W}$ is $CP$ conserving while the ${\cal O}_{BW}$, ${\cal O}_{WW}$  and ${\cal O}_{BB}$ are $CP$-violating operators.

In the SM, Dim-6 operators have no impact at the tree level on aNTGC, but at one-loop, there is an effect on aNTGC proportional to $\sigma_{dim6}$ $\sim$ ${\alpha \hat{s}}/{4\pi\Lambda^2}$. In the context of tree-level interactions, the effects of dim-8 operators become at the $\sigma_{dim8}$ $\sim$ ${\upsilon^2\hat{s}}/{\Lambda^4}$ level \cite{Grzadkowski:2010es,Buchmuller:1985jz}. The suppression of the dim-6 relative to the dim-8 contribution, expressed as $\frac{\sigma_{\text{dim8}}}{\sigma_{\text{dim6}}}$, can be described by the following equation.

\begin{eqnarray}
\Lambda \lesssim \sqrt{\frac{4\pi v^2}{\alpha}} \sim 10 \, \text{TeV}.
\end{eqnarray}

This condition surpass the one-loop effects originating from dim-6 operators \cite{Degrande:2014ydn}. The results presented in Table II, corresponding to 
$\Lambda\sim2$ TeV, indicate that the contributions arising from dim-6 terms are negligible within the context of this specific study. On the other hand, additively to the requirements of Lorentz and $U(1)_{em}$ invariance, Bose statistics introduce constraints, notably forbidding interactions vertices such as $ZZZ$, $ZZ\gamma$, or $Z\gamma\gamma$ when all particles are on-shell. The existence of such vertices necessitates the involvement of at least one off-shell gauge boson. The most comprehensive form entails only two independent couplings for each VZZ vertex (V = $\gamma$, Z; one CP-conserving and one CP-violating) and four independent couplings for each VZ$\gamma$ vertex (V = $\gamma$, Z; two CP-conserving and two CP-violating). Notably, no relationship exists between these couplings. Assuming these, the most general form of the $V_{1}V_{2}V_{3}$ vertex defined , where $V_{1},V_{2}$ are on-shell neutral gauge bosons, while $V_{3} = Z$ , $\gamma$ is in general off-shell. The effective Lagrangian for the neutral triple gauge couplings containing dim-6 and dim-8 operators is given in \cite{Gounaris:2000svs}.

\begin{eqnarray}
\label{eq.9}
\begin{split}
{\cal L}_{\text{aNTGC}}^{\text{dim-six,eight}}=&\frac{e}{m_Z^2}\Bigg[-[f_4^\gamma(\partial_\alpha F^{\alpha\beta})+f_4^Z(\partial_\alpha Z^{\alpha\beta})]Z_\mu (\partial^\mu Z_\beta)+[f_5^\gamma(\partial^\beta F_{\beta\mu})+f_5^Z (\partial^\beta Z_{\beta\nu})]\widetilde{Z}^{\nu\alpha}Z_\alpha  \\
&-[h_1^\gamma (\partial^\beta F_{\beta\mu})+h_1^Z (\partial^\alpha Z_{\alpha\mu})]Z_\beta F^{\mu\beta}-[h_3^\gamma(\partial_\beta F^{\beta\rho})+h_3^Z(\partial_\beta Z^{\beta\lambda})]Z^\sigma \widetilde{F}_{\lambda\sigma}   \\
&-\bigg\{\frac{h_2^\gamma}{m_Z^2}[\partial_\rho \partial_\beta \partial^\alpha F_{\alpha\nu}]+\frac{h_2^Z}{m_Z^2}[\partial_\rho\partial_\beta(\square+m_Z^2)Z_\nu]\bigg\}Z^\rho F^{\nu\beta}   \\
&+\bigg\{\frac{h_4^\gamma}{2m_Z^2}[\square\partial^\sigma F^{\rho\alpha}]+\frac{h_4^Z}{2m_Z^2}[(\square+m_Z^2)\partial^\rho Z^{\alpha\beta}]\bigg\}Z_\rho\widetilde{F}_{\alpha\beta}\Bigg].
\end{split}
\end{eqnarray}

Here, \( Z_{\alpha\beta} = \partial_\alpha Z_\beta - \partial_\beta Z_\alpha \), and \( \tilde{Z}_{\mu\nu} = \frac{1}{2}\varepsilon_{\mu\nu\rho\sigma}Z^{\rho\sigma} \) (with \( \varepsilon_{0123} = +1 \)) are the field strength tensor. Similarly, \( F_{\mu\nu} \) is the electromagnetic field tensor. However, \( f_4^V, h_1^V, h_2^V \) are three CP-violating, and \( f_5^V, h_3^V, h_4^V \) are three CP-conserving couplings (\( V = \gamma, Z \)). All couplings are zero at the tree-level in the SM. In the Lagrangian, the \( h_2^V \) and \( h_4^V \) couplings are dim-8, while the other four are dim-6.

The couplings in Eq. 10 are related to the coefficients given in Eqs. (2-5) under the $SU(2)_{L} \times U(1)_{Y}$ gauge invariance \cite{Rahaman:2020fdf}. The anomalous couplings for the \( ZZV \) interaction, involving two on-shell \( Z \)-bosons and one off-shell \( V = \gamma \) or \( Z \) boson with CP-conserving, are given below \cite{Degrande:2014ydn}:

\begin{eqnarray}
\label{eq.10}
f_5^Z=0,
\end{eqnarray}
\begin{eqnarray}
\label{eq.11}
f_5^\gamma=\frac{\upsilon^2 m_Z^2}{4c_\omega s_\omega} \frac{C_{\widetilde{B}W}}{\Lambda^4}.
\end{eqnarray}

{\raggedright However, the $CP$-violating couplings are the follows}

\begin{eqnarray}
\label{eq.12}
f_4^Z=\frac{m_Z^2 \upsilon^2 \left(c_\omega^2 \frac{C_{WW}}{\Lambda^4}+2c_\omega s_\omega \frac{C_{BW}}{\Lambda^4}+4s_\omega^2 \frac{C_{BB}}{\Lambda^4}\right)}{2c_\omega s_\omega},
\end{eqnarray}
\begin{eqnarray}
\label{eq.13}
f_4^\gamma=-\frac{m_Z^2 \upsilon^2 \left(-c_\omega s_\omega \frac{C_{WW}}{\Lambda^4}+\frac{C_{BW}}{\Lambda^4}(c_\omega^2-s_\omega^2)+4c_\omega s_\omega \frac{C_{BB}}{\Lambda^4}\right)}{4c_\omega s_\omega}.
\end{eqnarray}

In the equations below, $s_\omega$ and $c_\omega$ represent the sine and cosine of the weak mixing angles $\theta_{w}$. The $CP$-conserving coupling, featuring one on-shell $Z$-boson and a photon with one extra off-shell $V=Z$ or $\gamma$ boson, are given as follows \cite{Degrande:2014ydn}.

\begin{eqnarray}
\label{eq.14}
h_3^Z=\frac{\upsilon^2 m_Z^2}{4c_\omega s_\omega} \frac{C_{\widetilde{B}W}}{\Lambda^4},
\end{eqnarray}
\begin{eqnarray}
\label{eq.15}
h_4^Z=h_4^\gamma=h_3^\gamma=0.
\end{eqnarray}

{\raggedright Furthermore, the $CP$-violating couplings are characterized by the following expressions.

\begin{eqnarray}
\label{eq.16}
h_1^Z=\frac{m_Z^2 \upsilon^2 \left(-c_\omega s_\omega \frac{C_{WW}}{\Lambda^4}+\frac{C_{BW}}{\Lambda^4}(c_\omega^2-s_\omega^2)+4c_\omega s_\omega \frac{C_{BB}}{\Lambda^4}\right)}{4c_\omega s_\omega},
\end{eqnarray}
\begin{eqnarray}
\label{eq.17}
h_2^Z=h_2^\gamma=0,
\end{eqnarray}
\begin{eqnarray}
\label{eq.18}
h_1^\gamma=-\frac{m_Z^2 \upsilon^2 \left(s_\omega^2 \frac{C_{WW}}{\Lambda^4}-2c_\omega s_\omega \frac{C_{BW}}{\Lambda^4}+4c_\omega^2 \frac{C_{BB}}{\Lambda^4}\right)}{4c_\omega s_\omega}.
\end{eqnarray}

The connections of $C_{WW}/\Lambda^4$, $C_{BW}/\Lambda^4$, $C_{BB}/\Lambda^4$, and $C_{\tilde{B}W}/\Lambda^4$ in Equations (12)-(15), (17), (19) define the dim-8 aNTGC that CP-conserving $C_{\tilde{B}W}/\Lambda^4$ and CP-violating $C_{WW}/\Lambda^4$, $C_{BB}/\Lambda^4$, $C_{BW}/\Lambda^4$ couplings. 

On the other hand, dimension-8 operators naturally emerge in scenarios involving new physics sectors, particularly with strong interactions. Such operators can arise at loop level, for instance, from the contributions of heavy fermions or scalar multiplets. In this context, fermions with an axial coupling to the Z-boson can induce CP-conserving operator through one-loop effects, generating terms proportional to the Levi-Civita tensor\cite{Degrande:2013kka}. These contributions help explain the presence of operators like $C_{\tilde{B}W}/\Lambda^4$ given in the following equation

\begin{equation}
\frac{C_{\Tilde{B}W}}{\Lambda^4} = \frac{e^2 Q g_A g_V}{2 \pi^2 c_W s_W M_F^2},
\end{equation}

where $Q$, $g_A$, and $g_V$ are the fermion charge, axial, and vector couplings to the $Z$-boson, respectively, and $M_F$ is the mass of the fermion. This relationship demonstrates that dimension-8 operators can arise through loop effects, with coefficients inversely proportional to the square of the fermion mass.

The experimental results on dim-8 couplings are determined through the $pp \rightarrow Z\gamma \rightarrow \nu\bar{\nu}\gamma$ process, including neutrino decay, at LHC with a luminosity of 36.1 fb$^{-1}$ at $\sqrt{s}=13$ TeV \cite{Aaboud:2018ybz}. The experimental sensitivities in this study, at a 95\% Confidence Level (C.L.), are provided as follows:

\begin{eqnarray}
\label{eq.19}
-1.1\, \text{TeV}^{-4}<\frac{C_{\widetilde{B}W}}{\Lambda^4}<1.1 \, \text{TeV}^{-4},
\end{eqnarray}
\begin{eqnarray}
\label{eq.20}
-2.3\, \text{TeV}^{-4}<\frac{C_{WW}}{\Lambda^4}<2.3 \, \text{TeV}^{-4},
\end{eqnarray}
\begin{eqnarray}
\label{eq.21}
-0.65\, \text{TeV}^{-4}<\frac{C_{BW}}{\Lambda^4}<0.64 \, \text{TeV}^{-4},
\end{eqnarray}
\begin{eqnarray}
\label{eq.22}
-0.24\, \text{TeV}^{-4}<\frac{C_{BB}}{\Lambda^4}<0.24 \, \text{TeV}^{-4}.
\end{eqnarray}

\section{Event Generation and Cut-Based Analysis}

The investigation of the $C_{\tilde{B}W}/\Lambda^4$, $C_{BW}/\Lambda^4$, $C_{BB}/\Lambda^4$, and $C_{WW}/\Lambda^4$ couplings, which affect the $ZZ\gamma$ and $Z\gamma\gamma$ couplings within the $\mu^- \gamma \to Z l^{-} \to l^{-} \tilde{\nu_{l}} \nu_{l}$ process at the FCC and SPPC, operating at $\sqrt{s}=24.5$ and 20.2 TeV with integrated luminosities of ${\cal L}_{int}=5$fb$^{-1}$ and 42.8 fb$^{-1}$, respectively. 
The analysis was performed within the Effective Field Theory (EFT) framework, focusing on the dimension-8 operators that contribute to anomalous triple gauge couplings. We employed the aNTGC model implemented in the MadGraph5aMC@NLO. The parameter space was explored by varying the Wilson coefficients of the relevant operators while keeping other EFT parameters fixed.
Both signal-background analyses for the $\mu^- \gamma \to Z l^{-} \to l^{-} \tilde{\nu_{l}} \nu_{l}$ process are produced, incorporating the aNTGCs via the Universal Feynrules Output(UFO) into MadGraph5aMC@NLO \cite{Alwall:2014cvc,Degrande:2013kka}. The study focuses on the $\nu\nu\ell$ final state, presuming the $Z$-boson decays into a neutrino pair during $Z\ell$ production, where $\ell$ denotes charged leptons ($\ell=e,\mu$). Processes involving $Z$-boson decay into neutrinos offer distinct advantages over those involving decay into charged leptons ($\ell^{-}\ell^{+}$) or hadrons ($q\bar{q}$). Hadronic decay leads to challenges in obtaining clean data due to significant QCD backgrounds. In contrast, neutrino pair decay provides a higher branching ratio for the $Z$-boson, offering enhanced sensitivity in higher-energy regions.

This study based the EPA, also known as the Weizsacker Williams Approximation (WWA) \cite{Weizsacker,Williams}, as the primary theoretical framework for evaluating the sensitivity of the $\mu^- \gamma \to Z l^{-} \to l^{-} \tilde{\nu_{l}} \nu_{l}$ process. Here, the emitted $\gamma$ is a quasi-real photon scattered from the incoming proton beam according to the framework of the Weizsacker-Williams Approximation (WWA). This quasi-real photon is scattered at small angles from the beam pipe. These photons have low virtuality, and they can almost be supposed to be on mass-shell. Hence, they are assumed to be almost real. The analysis aims to evaluate the potential of FCC and SPPC muon-proton colliders in probing anomalous $Z\gamma\gamma$ and $ZZ\gamma$ triple vertices. The subsequent equation illustrates the photon distribution emitted by a proton beam \cite{Budnev,Chen2}.

\begin{eqnarray}
f_{\gamma(x)}=\frac{\alpha}{\pi E_{p}}\{[1-x][\varphi(\frac{Q_{max}^{2}}{Q_{0}^{2}})-\varphi(\frac{Q_{min}^{2}}{Q_{0}^{2}})],
\end{eqnarray}

\noindent where the function $\varphi$ is given as:

\begin{eqnarray}
\varphi(\theta)=&&(1+ky)\left[-\textit{In}(1+\frac{1}{\theta})+\sum_{s=1}^{3}\frac{1}{s(1+\theta)^{s}}\right]+\frac{y(1-l)}{4\theta(1+\theta)^{3}} \nonumber \\
&& +m(1+\frac{y}{4})\left[\textit{In}\left(\frac{1-l+\theta}{1+\theta}\right)+\sum_{s=1}^{3}\frac{b^{s}}{s(1+\theta)^{s}}\right]. \nonumber \\
\end{eqnarray}

\noindent Here,

\begin{eqnarray}
y=\frac{x^{2}}{(1-x)},
\end{eqnarray}

\begin{eqnarray}
k=\frac{1+\mu_{p}^{2}}{4}+\frac{4m_{p}^{2}}{Q_{0}^{2}}\approx 7.16,
\end{eqnarray}

\begin{eqnarray}
l=1-\frac{4m_{p}^{2}}{Q_{0}^{2}}\approx -3.96,
\end{eqnarray}

\begin{eqnarray}
m=\frac{\mu_{p}^{2}-1}{b^{4}}\approx 0.028.
\end{eqnarray}

The cross-section of the total process $\mu^- \gamma \to Z l^{-} \to l^{-} \tilde{\nu_{l}} \nu_{l}$  is thus obtained from:

\begin{eqnarray}
\sigma=\int f_{\gamma}(x) d\hat{\sigma}dE_{1}.
\end{eqnarray}

The Feynman diagrams for the $\mu^- \gamma \to Z l^{-} \to l^{-} \tilde{\nu_{l}} \nu_{l}$ process are given in Figure 1. Here, the two diagrams on the bottom include contributions from the Standard Model, while the two Feynman diagrams on the top include those coming beyond the Standard Model, associated with anomalous $Z\gamma\gamma$ and $ZZ\gamma$ vertices.

\begin{figure}[H]
\centerline{\scalebox{0.3}{\includegraphics{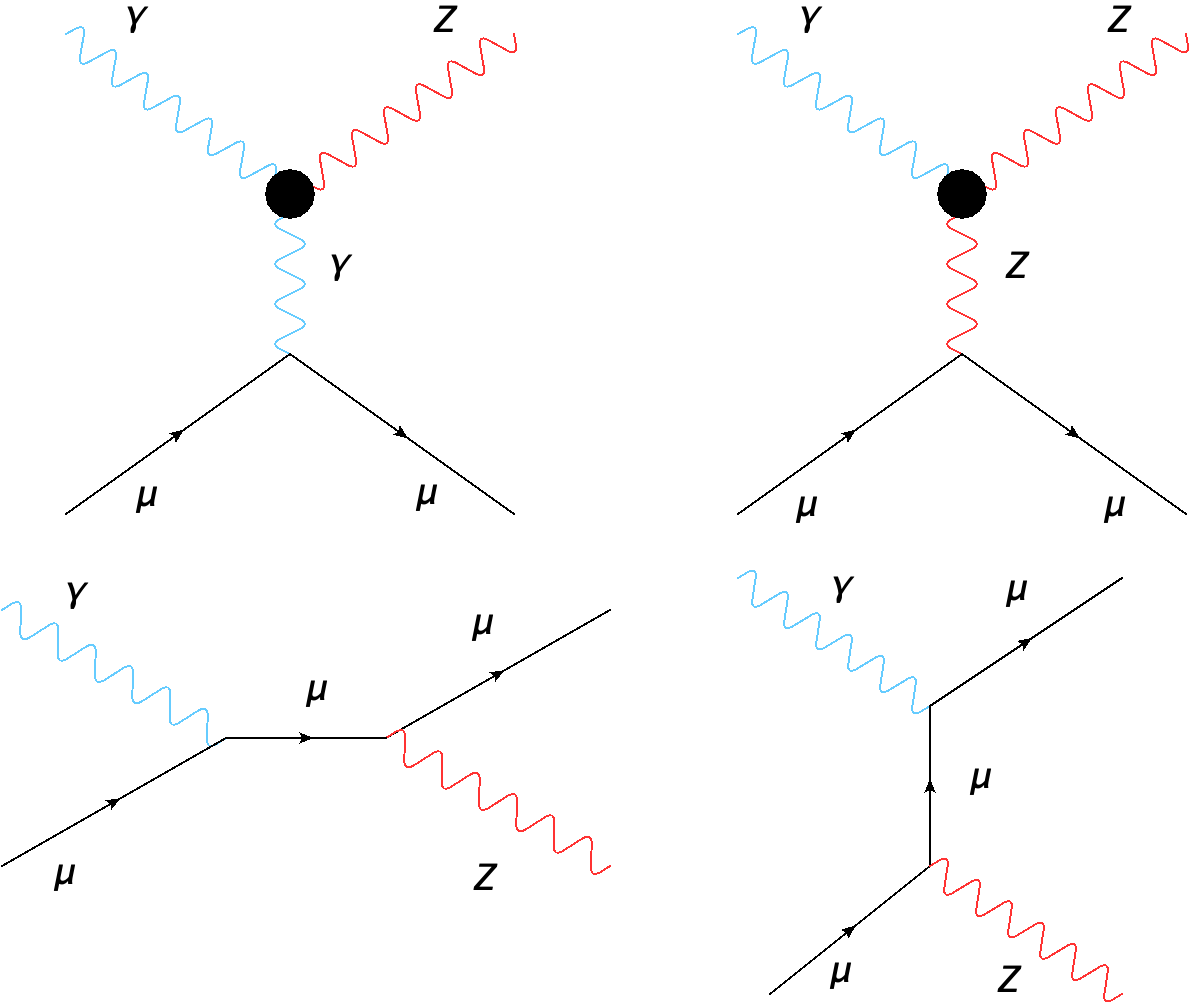}}}
\caption{ \label{fig:gamma}  Diagrams for the process $\mu^- \gamma \to Z l^{-}$
including the anomalous $ZZ\gamma$ and $Z\gamma\gamma$ couplings. New physics contributions are shown by a black circle. Rest of them are contain the Standard Model contributions.}
\end{figure}

In the study, kinematic cuts are applied to separate the signal from the background. Specifically, cuts on the transverse momentum $p^{T}$ and pseudo-rapidity $\eta$ of the charged lepton, as well as the missing transverse energy $\slashed{E}_T$. These cuts effectively suppress the relevant background and enhance the signal-to-background ratio. We obtained the optimized cuts with the help of the kinematic distributions of the variables given in Fig. 2-4. 

\begin{figure}{H}
\centering
\begin{subfigure}[t]{1.05\textwidth}
   \includegraphics[width=1\linewidth]{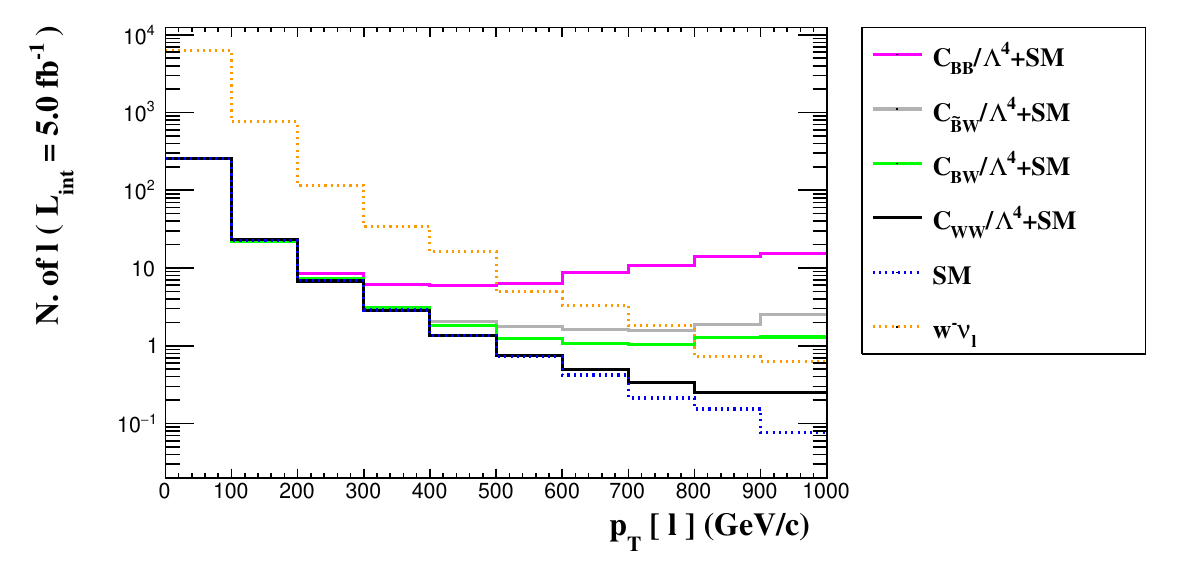}
   \caption{}
   \label{Figure2} 
\end{subfigure}

\begin{subfigure}[b]{1.05\textwidth}
   \includegraphics[width=1\linewidth]{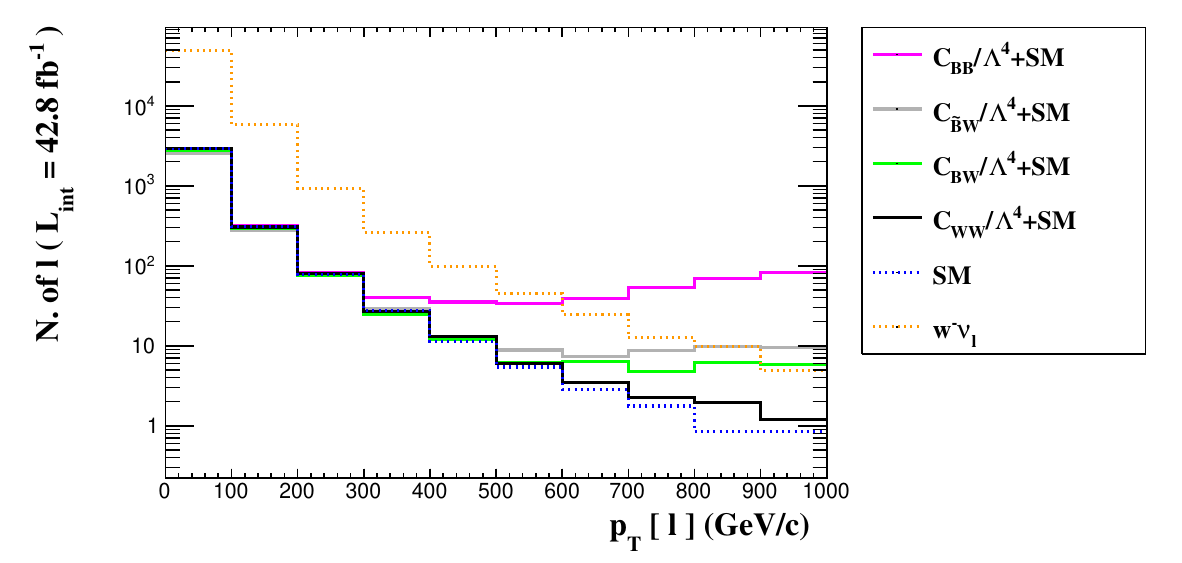}
   \caption{}
   \label{Figure3}
\end{subfigure}

\caption[]{(a) The number of events as a function of the $p^{\ell}_T$ for the process $\mu^- \gamma \to Z l^{-} \to l^{-} \tilde{\nu_{l}} \nu_{l}$ and related backgrounds at FCC with the $\sqrt{s}=24.5$ TeV. (b) Same as for (a) but for SPPC with the $\sqrt{s}=20.2$ TeV.}
\end{figure}

\begin{figure}{H}
\centering
\begin{subfigure}[b]{1.05\textwidth}
   \includegraphics[width=1\linewidth]{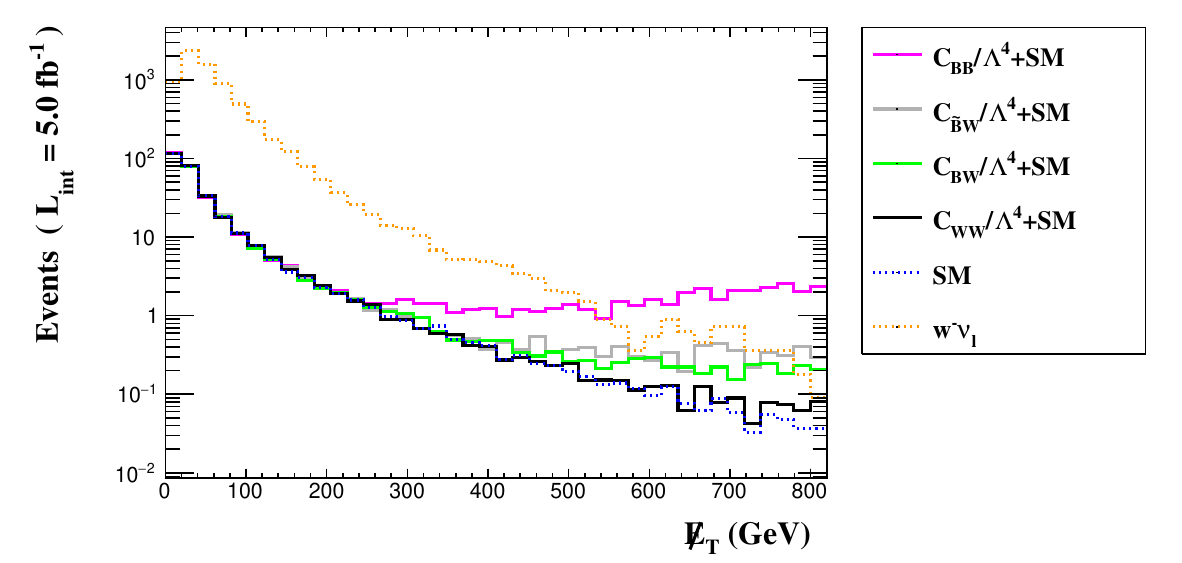}
   \caption{}
   \label{Figure4} 
\end{subfigure}

\begin{subfigure}[b]{1.05\textwidth}
   \includegraphics[width=1\linewidth]{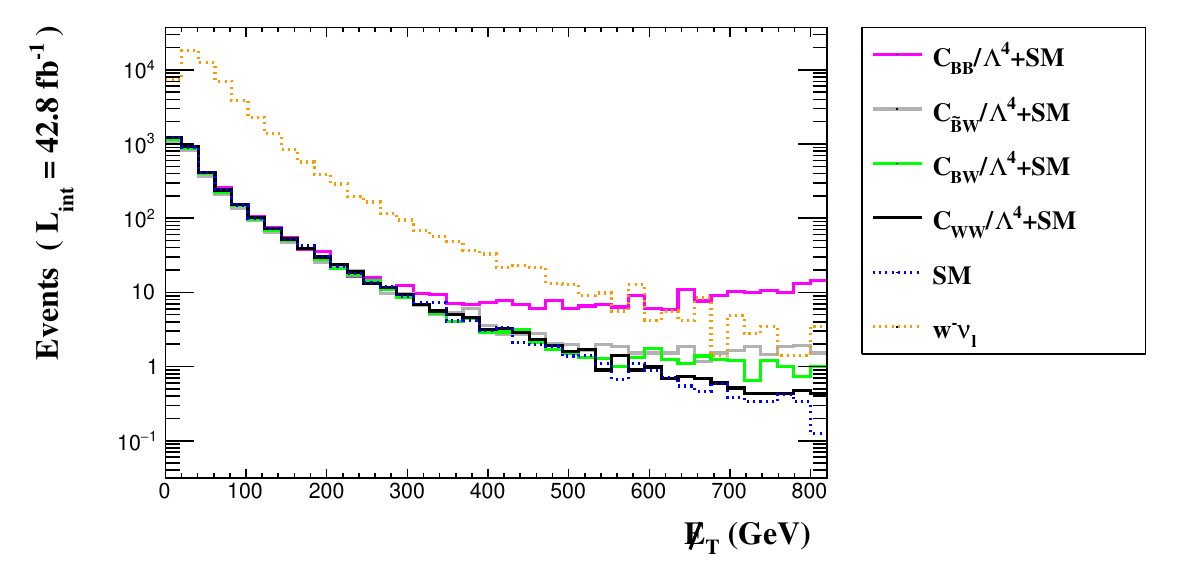}
   \caption{}
   \label{Figure5}
\end{subfigure}

\caption[]{(a) The number of events as a function of the $\slashed{E}_T$ for the process $\mu^- \gamma \to Z l^{-} \to l^{-} \tilde{\nu_{l}} \nu_{l}$ and related backgrounds at FCC with the $\sqrt{s}=24.5$ TeV. (b) Same as for (a) but for SPPC with the $\sqrt{s}=20.2$ TeV.}
\end{figure}

\begin{figure}{H}
\centering
\begin{subfigure}[b]{1.05\textwidth}
   \includegraphics[width=1\linewidth]{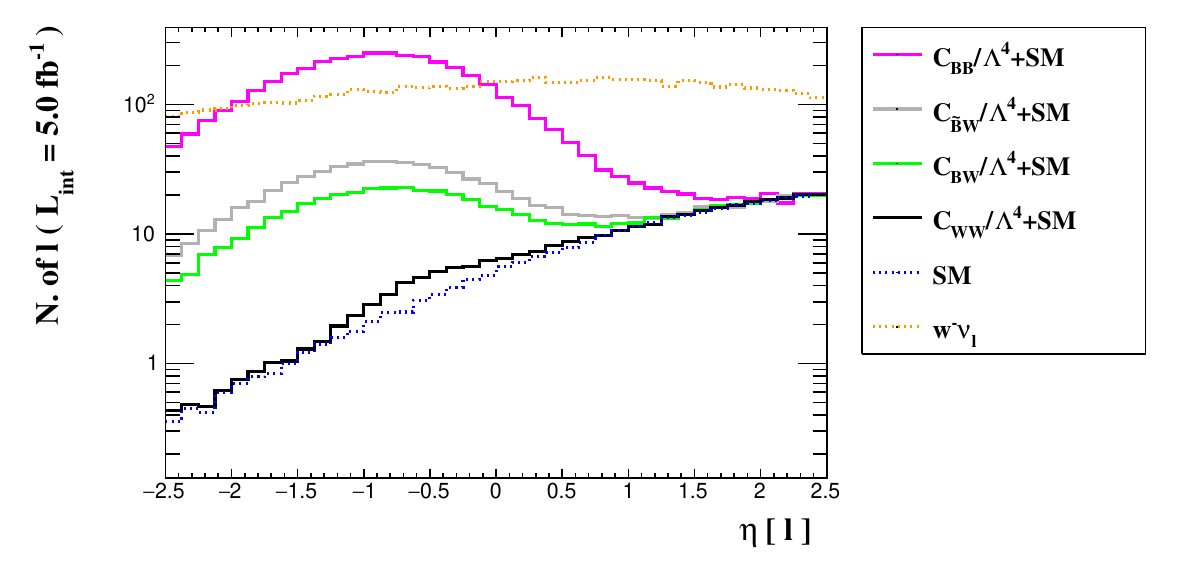}
   \caption{}
   \label{Figure6} 
\end{subfigure}

\begin{subfigure}[b]{1.05\textwidth}
   \includegraphics[width=1\linewidth]{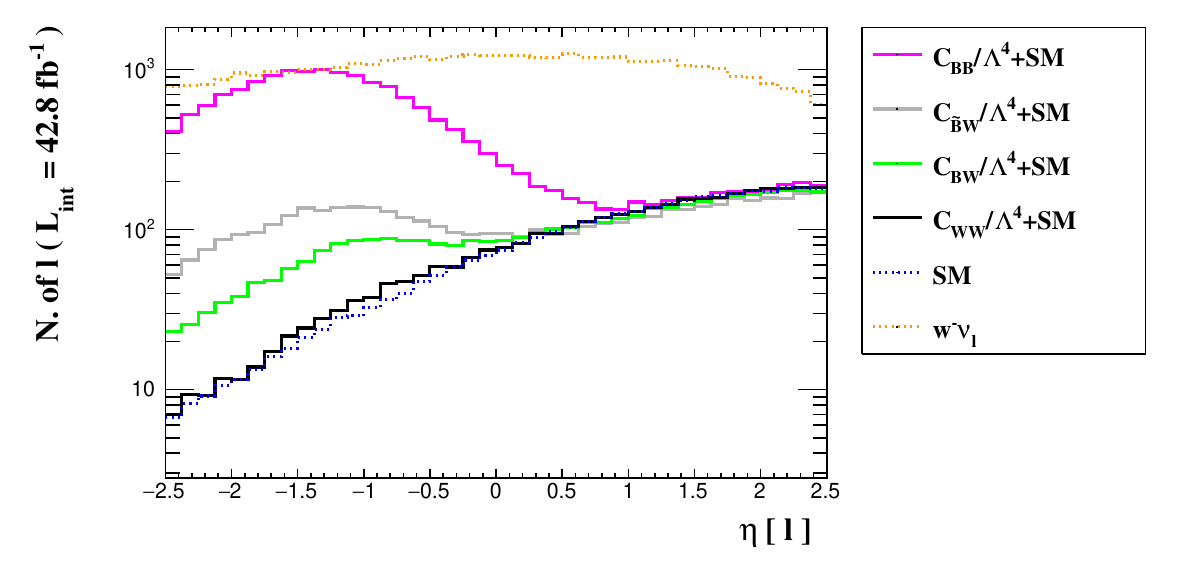}
   \caption{}
   \label{Figure7}
\end{subfigure}

\caption[]{(a) The number of events as a function of $\eta_{\ell}$ for the process $\mu^- \gamma \to Z l^{-} \to l^{-} \tilde{\nu_{l}} \nu_{l}$ and related backgrounds at FCC with the $\sqrt{s}=24.5$ TeV. (b) Same as for (a) but for SPPC with the $\sqrt{s}=20.2$ TeV.}
\end{figure}

Beholding the separation point of the signals and backgrounds in the related figures and the number of events, we obtained the transverse momentum and the pseudo-rapidity of the charged lepton $p^{l}_T>80$ GeV, $\eta_{l}<2$ and the transverse missing energy $\slashed{E}_T > 300$ GeV, based on the final state particles of the process $\mu^- \gamma \to Z l^{-} \to l^{-} \tilde{\nu_{l}} \nu_{l}$. The number of events after the applied cuts is given step-by-step in Table-I for the signals and the related backgrounds at FCC-$\mu p$ and SPPC-$\mu p$.

\begin{table}{H}
\caption{Events for the process $\mu^- \gamma \to Z l^{-} \to l^{-} \tilde{\nu_{l}} \nu_{l}$ for various couplings and SM background. Here, $\epsilon[\%]$ is the relative efficiency of each cut.}
\begin{tabular}{|c|c|c|c|c|c|c|}
\hline
\multicolumn{3}{|c|} {Background} & \multicolumn{4}{|c|} {Signals} \\
\hline
Cuts&Standard Model & $W^{-}\nu_{l}\to l^{-} \tilde{\nu_{l}} \nu_{l}$& $C_{BB}/\Lambda^{4}$ & $C_{\tilde{B}W}/\Lambda^{4}$ & $C_{BW}/\Lambda^{4}$ &$C_{WW}/\Lambda^{4}$ \\
\hline
FCC-$\mu p$ & Events \, $\epsilon[\%]$ & Events \, $\epsilon[\%]$ & Events \, $\epsilon[\%]$ & Events \, $\epsilon[\%]$ & Events \, $\epsilon[\%]$ &  Events \, $\epsilon[\%]$ \\
\hline
Presel. & 293 \, \, \, --- & 7200 \, \, \, --- & 4131 \, \, \, --- & 830 \,\,\, --- &613 \,\,\, ---& 312 \,\,\, --- \\

$p^l_T > 80$ GeV  & 47 \, \, \, 16.0 & 1451 \, \, \, 20.2 & 3883 \, \, \, 94.0 & 584 \, \, \, 70.3 & 368 \, \, \, 60.0 & 65 \, \, \, 20.9   \\

$\slashed{E}_T > 300$ GeV  & 6 \, \, \, 12.8 & 64 \, \, \, 4.4 & 3841 \, \, \, 99.0 & 542 \, \, \, 92.8 & 328 \, \, \, 88.9 & 23 \, \, \, 36.1 \\

$\eta^{l} < 2.0$  & 5 \, \, \, 75.6 & 63 \, \, \, 98.6 & 3838 \, \, \, 99.9 & 540 \, \, \, 99.7 & 326 \, \, \, 99.4 & 22 \, \, \, 94.9 \\
\hline
SPPC-$\mu p$ &  &  &  &  &  &   \\
\hline
Presel. & 3360 \, \, \, --- & 55982 \, \,  --- & 17483 \, \, \,  --- & 4682 \,\,\, --- & 3968 \,\,\, ---& 3429 \,\,\, --- \\

$p^l_T > 100$ GeV  & 439 \, \, \, 13.1 & 7168 \, \, \, 12.8 & 14546 \, \, \, 83.1 & 2109 \, \, \, 34.0 & 1196 \, \, \, 30.1 & 508 \, \, \, 14.8   \\

$\slashed{E}_T > 400$ GeV  & 24 \, \, \, 5.5 & 205 \, \, \, 2.9 & 14108 \, \, \, 96.9 & 1727 \, \, \, 81.8 & 805 \, \, \, 67.2 & 96 \, \, \, 18.8 \\

$\eta^{l} < 2.0$  & 24 \, \, \, 100.0 & 205 \, \, \, 100.0 & 14108 \, \, \, 100.0 & 1727 \, \, \, 100.0 & 805 \, \, \, 100.0 & 96 \, \, \, 100.0 \\
\hline
\end{tabular}
\end{table}

Here, all couplings are chosen 1 TeV$^{-4}$. After the applied cuts, the SM and the $W^{-}\nu_{l} \to l^{-} \tilde{\nu_{l}} \nu_{l}$ background events decreased dramatically compering with the preselection case. Truncated background events allow more confident analysis to get the sensitivities on the dim-8 aNTGCs. The coefficients of dimension-eight operators can be connected to the new physics characteristic scale $\Lambda$, and an upper bound can be imposed on this scale. For C = O(1) couplings, we calculate $\Lambda < \sqrt{4\pi\upsilon\sqrt{\hat{s}}}$ $\sim$ 4.9 TeV for FCC and 4.47 TeV for SPPC. On the other hand, the EFT is not a complete model and violates unitarity at sufficiently high energy scales. The form factor scheme is a scheme where the effects ultimately depend on the value of the cut-off scale. The standard procedure to avoid this unphysical behavior of the cross-section and to obtain meaningful limits is to multiply the anomalous couplings by a dipole form factor of the form \cite{Ellison-ARNPS1998,Eboli4}:

\begin{equation}
FF=\frac{1}{(1+\frac{\hat s}{\Lambda^2_{FF}})^2}.
\end{equation}

\noindent In Eq. (32), $\hat s$ corresponds to the partonic center-of-mass energy, and $\Lambda_{FF}$ is the cut-off scale of the dipole form factor, which corresponds to the energy regime above which the contributions of the anomalous couplings are largely suppressed. In this study, we have used the $\Lambda_{FF}=3$ TeV to handle the unitarity violation.

On the other hand, we give the total cross-sections of the process $\mu^- \gamma \to Z l^{-} \to l^{-} \tilde{\nu_{l}} \nu_{l}$ in terms of the anomalous $C_{BB}/\Lambda^4$, $C_{BW}/\Lambda^4$, $C_{WW}/\Lambda^4$, and $C_{\tilde{B}W}/\Lambda^4$ couplings at FCC and SPPC in Fig. 5-6. In these figures, each coupling is set individually. Here, only one of the aNTGC couplings is non-zero at any time, while the others are zero. These figures show the deviation of the couplings from the SM. In these figures, $C_{BB}/\Lambda^4$ is more significant than the other couplings at both FCC and SPPC.

\begin{figure}[H]
\centerline{\scalebox{1.45}{\includegraphics{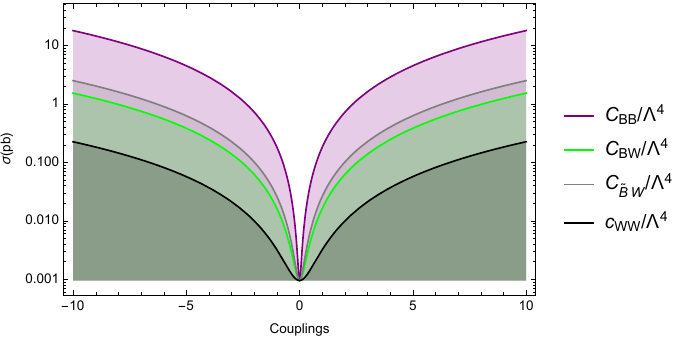}}}
\caption{ \label{fig:gamma} Cross-sections of the process $\mu^- \gamma \to Z l^{-} \to l^{-} \tilde{\nu_{l}} \nu_{l}$ in terms of the parameters  $C_{WW}/\Lambda^{4}$, $C_{BB}/\Lambda^{4}$, $C_{\widetilde{B}W}/\Lambda^{4}$, $C_{BW}/\Lambda^{4}$ at SPPC. }
\end{figure}

\begin{figure}[H]
\centerline{\scalebox{1.45}{\includegraphics{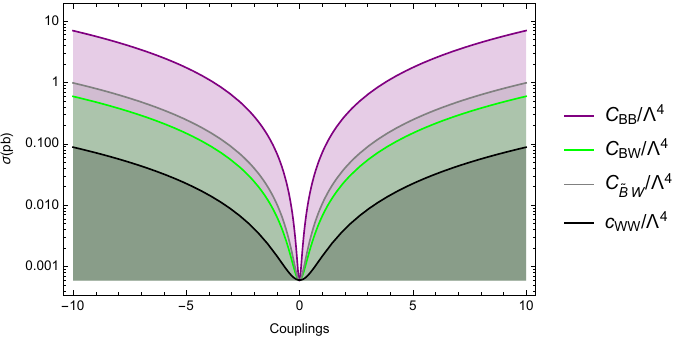}}}
\caption{ \label{fig:gamma} Cross-sections of the process $\mu^- \gamma \to Z l^{-} \to l^{-} \tilde{\nu_{l}} \nu_{l}$ in terms of the parameters $C_{WW}/\Lambda^{4}$, $C_{BB}/\Lambda^{4}$,  $C_{\widetilde{B}W}/\Lambda^{4}$, $C_{BW}/\Lambda^{4}$ at FCC.}
\end{figure}

\section{Expected Sensitivities on Dim-8 aNTGC}

To investigate the sensitivities of the anomalous $C_{BB}/\Lambda^4$, $C_{BW}/\Lambda^4$, $C_{WW}/\Lambda^4$, and $C_{\tilde{B}W}/\Lambda^4$ couplings at \%95 C.L., a $\chi^{2}$ test with systematic errors was applied. The $\chi^{2}$ test is defined as follows:

\begin{equation}
\chi^2(f/\Lambda^4)=\Biggl(\frac{\sigma_{SM}(\sqrt{s})-\sigma_{Total}(\sqrt{s}, f/\Lambda^4)}
{\sigma_{SM}(\sqrt{s})\sqrt{(\delta_{sys})^2 + (\delta_{st})^2}}\Biggr)^2,
\end{equation}

Here, $\sigma_{SM}(\sqrt{s})$ represents the cross-section of the SM background, and $\sigma_{total}(\sqrt{s}, f/\Lambda^4)$ is the total cross-section of both the new physics coming from beyond the SM and the SM background.  $\delta_{st}=\frac{1}{\sqrt{N_{SM}}}$ and $\delta_{sys}$ are the statistical error and systematic uncertainty, respectively. The event number of the SM background is defined as $N_{SM}={\cal L}\times \sigma_{SM}$, where ${\cal L}$ is the integrated luminosity. Systematic uncertainties from various sources have been included in the $\chi^{2}$ analysis \cite{khoriauli}. Possible sources of systematic uncertainties are integrated luminosities, photon efficiencies, jet-photon misidentification, detector efficiency and background estimation. In the study, systematic uncertainties of 0\% and 5\% have been considered. Obtained sensitivities on aNTGs via the process $\mu^- \gamma \to Z l^{-} \to l^{-} \tilde{\nu_{l}} \nu_{l}$ at FCC-$\mu p$ and SPPC-$\mu p$ under systematic uncertainties of 0\% and 5\% for $\Lambda_{FF}= \infty$ and 3 TeV are given in Table II, respectively. On the other hand, Figs. 7-8 compare the obtained sensitivities at FCC-$\mu$ p and SPPC-$\mu$ p  and the experimental results for $\Lambda_{FF}= \infty$ and 3 TeV.

\begin{table}[H]
\centering
\caption{Sensitivities at $95\%$ C.L. on the anomalous $Z\gamma\gamma$ and $ZZ\gamma$ couplings for different $\Lambda$ cut-off values and various systematic uncertainties.}
\begin{tabular}{|c|c|c|c|c|c|}
\hline
\multicolumn{4}{|c|} {} & \multicolumn{1}{|c|} {FCC-$\mu p$} & \multicolumn{1}{|c|} {SPPC-$\mu p$} \\
\hline
Couplings & ATLAS \cite{Aaboud:2018ybz} & CMS \cite{CMS:2020gtj} & Sys Errors & Our Projection & Our Projection \\
\hline
\multicolumn{6}{|c|}{\textbf{$\Lambda_{FF} =\infty$}} \\
\hline
& & & $\delta=0\%$ & $[-0.068; 0.070]$ & $[-0.066; 0.051]$  \\
$C_{BB}/\Lambda^{4}$  & $[-0.24; 0.24]$ & $[-1.20; 1.20]$ & $\delta=5\%$ & $[-0.069; 0.070]$ & $[-0.067; 0.051]$ \\
\hline
& & & $\delta=0\%$ & $[-0.236; 0.238]$ & $[-0.207; 0.192]$  \\
$C_{BW}/\Lambda^{4}$  &$[-0.65; 0.64]$ & $[-1.40; 1.30]$ & $\delta=5\%$ & $[-0.237; 0.239]$ & $[-0.210; 0.195]$  \\
\hline
& & & $\delta=0\%$ & $[-0.187; 0.184]$ & $[-0.158; 0.151]$  \\
$C_{\widetilde{B}W}/\Lambda^{4}$  &  $[-1.10; 1.10]$ & $[-2.30; 2.50]$ & $\delta=5\%$ & $[-0.187; 0.185]$ & $[-0.160; 0.154]$  \\
\hline
& & & $\delta=0\%$ & $[-0.619; 0.621]$ & $[-0.514; 0.522]$  \\
$C_{WW}/\Lambda^{4}$  & $[-2.30; 2.30]$ & $[-1.40; 1.20]$ & $\delta=5\%$ & $[-0.621; 0.623]$ & $[-0.522; 0.530]$  \\
\hline
\multicolumn{6}{|c|}{\textbf{$\Lambda_{FF} = 3$ TeV}} \\
\hline
& & & $\delta=0\%$ & $[-0.198; 0.200]$ & $[-0.157; 0.151]$  \\
$C_{BB}/\Lambda^{4}$  & $[-0.24; 0.24]$ & $[-1.20; 1.20]$ & $\delta=5\%$ & $[-0.199; 0.200]$ & $[-0.159; 0.153]$ \\
\hline
& & & $\delta=0\%$ & $[-0.628; 0.629]$ & $[-0.489; 0.478]$  \\
$C_{BW}/\Lambda^{4}$  &$[-0.65; 0.64]$ & $[-1.40; 1.30]$ & $\delta=5\%$ & $[-0.629; 0.630]$ & $[-0.496; 0.485]$  \\
\hline
& & & $\delta=0\%$ & $[-0.187; 0.184]$ & $[-0.158; 0.152]$  \\
$C_{\widetilde{B}W}/\Lambda^{4}$  &  $[-1.10; 1.10]$ & $[-2.30; 2.50]$ & $\delta=5\%$ & $[-0.187; 0.185]$ & $[-0.160; 0.154]$  \\
\hline
& & & $\delta=0\%$ & $[-0.859; 0.860]$ & $[-0.704; 0.716]$  \\
$C_{WW}/\Lambda^{4}$  & $[-2.30; 2.30]$ & $[-1.40; 1.20]$ & $\delta=5\%$ & $[-0.862; 0.862]$ & $[-0.715; 0.728]$  \\
\hline
\end{tabular}
\end{table}

\begin{figure}[H]
\centerline{\scalebox{0.8}{\includegraphics{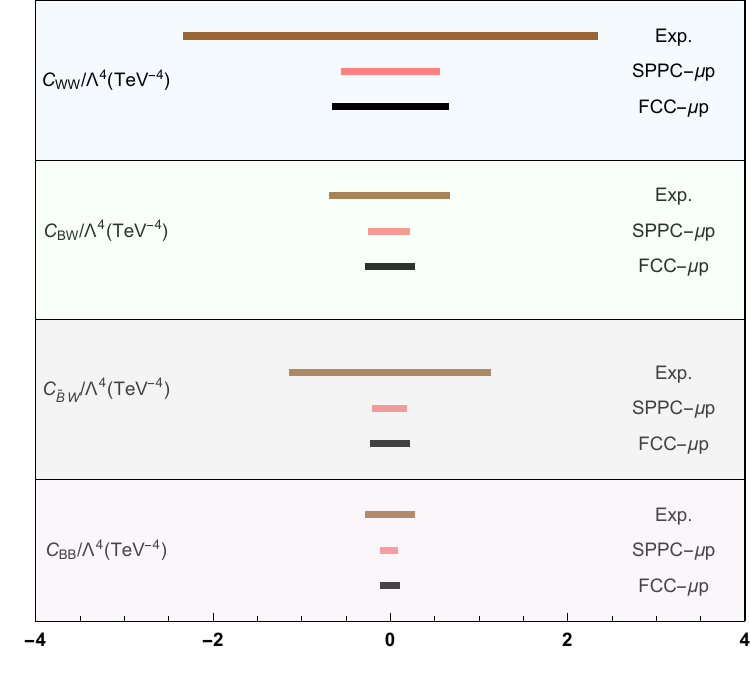}}}
\caption{ \label{fig:gamma} Comparison of LHC bounds and obtained sensitivities on the anomalous $C_{WW}/\Lambda^{4}$, $C_{BB}/\Lambda^{4}$, $C_{\widetilde{B}W}/\Lambda^{4}$, $C_{BW}/\Lambda^{4}$ parameters via the process $\mu^- \gamma \to Z l^{-} \to l^{-} \tilde{\nu_{l}} \nu_{l}$ for $\Lambda_{FF}=\infty$ at FCC and SPPC.}
\end{figure}

\begin{figure}[H]
\centerline{\scalebox{1.13}{\includegraphics{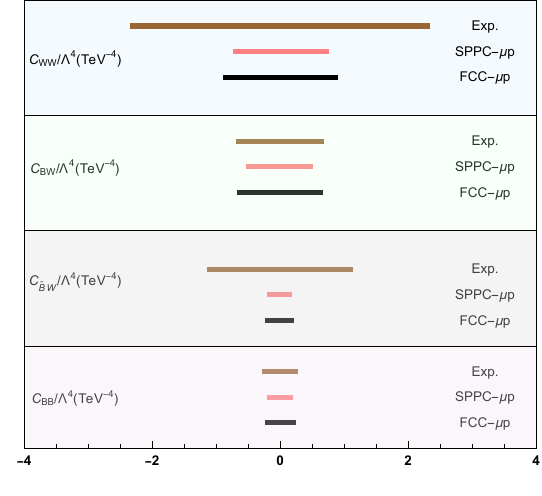}}}
\caption{ \label{fig:gamma} Same for Fig.7 but for $\Lambda_{FF}=3$ TeV }
\end{figure}

\section{Conclusions}

This study introduces a phenomenological approach using cut-based techniques to explore the constraints on the CP-conserving $C_{\tilde{B}W}/\Lambda^{4}$ coupling and CP-violating $C_{\tilde{B}W}/\Lambda^{4}$, $C_{WW}/\Lambda^{4}$ and $C_{BB}/\Lambda^{4}$ couplings via the process $\mu^- \gamma \to Z l^{-} \to l^{-} \tilde{\nu_{l}} \nu_{l}$ at the FCC and SPPC. It emphasizes the importance of $p^{l}_T$, $\eta_{l}$, and the transverse missing energy $\slashed{E}_T$ in distinguishing signals from the relevant background. By using the cut-based analysis, we determine the sensitivities of each anomalous coupling using the $\chi^{2}$ method for the FCC-$\mu p$ and SPPC-$\mu p$ colliders operating at $\sqrt{s}=24.5$ and 20.2 TeV with ${\cal L}_{int}=5$ and 42.8  fb$^{-1}$, respectively. Comparisons are made with the experimental research on $\nu\nu\gamma$ production at the LHC running at $\sqrt{s}=13$ TeV, as well as theoretical studies concerning $\nu\nu\gamma$ production at the HL-LHC and HE-LHC, and $Z\gamma\to\nu\nu\gamma$ production at the muon collider \cite{Senol:2020hbh,emre}.

While our results outperform those of the HL-LHC and the HE-LHC in general, they are  2-3 times worse than the sensitivities achieved by the muon collider. Consequently, our findings suggest that the $\mu p$ collisions at FCC and SPPC with $\sqrt{s}$=24.5 and 20.2 TeV could enhance the sensitivities on $C_{BB}/\Lambda^{4}$, $C_{\tilde{B}W}/\Lambda^{4}$, $C_{BW}/\Lambda^{4}$, $C_{WW}/\Lambda^{4}$ parameters defining the aNTGC $ZZ\gamma$ and $Z\gamma\gamma$ compared to the latest experimental results of LHC and the phenomenological results of future hadron-hadron colliders.

\begin{center}
{\bf Acknowledgements}
\end{center}

The numerical calculations reported in this paper were fully performed at TUBITAK ULAKBIM, High Performance and Grid Computing Center (TRUBA resources).
\vspace{1cm}

\section{Data Availability Statement}

This manuscript has no associated data or the data will not be deposited. [Authors’ comment: Data will be made available upon reasonable request.]


\begin{thebibliography}{99}

\bibitem{Baur:2000hfg}
U.~Baur and D.~Rainwater,
Phys.\ Rev.\ D {\bf 62}, 113011 (2000)
[arXiv:hep-ph/0008063].

\bibitem{Spor:2020wft}
S.~Spor and M.~K\"oksal,
Phys. Lett. B \textbf{820} (2021), 136533
doi:10.1016/j.physletb.2021.136533
[arXiv:2009.05848 [hep-ph]].

\bibitem{Billur:2019cav}
A.~A.~Billur, M.~K\"oksal, A.~Guti\'errez-Rodr\'\i{}guez and M.~A.~Hern\'andez-Ru\'\i{}z,
Eur. Phys. J. Plus \textbf{136} (2021) no.6, 697
doi:10.1140/epjp/s13360-021-01684-6
[arXiv:1909.10299 [hep-ph]].

\bibitem{Ari:2015tca}
V.~Ar\i{}, A.~A.~Billur, S.~C.~\.Inan and M.~K\"oksal,
Nucl. Phys. B \textbf{906} (2016), 211-230
doi:10.1016/j.nuclphysb.2016.02.029
[arXiv:1506.08998 [hep-ph]].

\bibitem{Choudhury:1994ywq}
D.~Choudhury and S.~D.~Rindani,
Phys.\ Lett.\ B {\bf 335}, 198-204 (1994)
[arXiv:hep-ph/9405242].

\bibitem{Atag:2004ybz}
S.~Ata\u{g} and \.{I}.~\c{S}ahin,
Phys.\ Rev.\ D {\bf 70}, 053014 (2004)
[arXiv:hep-ph/0408163].

\bibitem{Ots:2004twm}
I.~Ots, H.~Uibo, H.~Liivat, R.~K.~Loide and R.~Saar,
Nucl.\ Phys.\ B {\bf 702}, 346-356 (2004).

\bibitem{Ots:2006gsd}
I.~Ots, H.~Uibo, H.~Liivat, R.~K.~Loide and R.~Saar,
Nucl.\ Phys.\ B {\bf 740}, 212-221 (2006).

\bibitem{Rodriguez:2009rnw}
A.~Guti{\'e}rrez-Rodr{\'{\i}}guez, M.~A.~Hern{\'a}ndez-Ru{\'{\i}}z and M.~A.~P{\'e}rez,
Phys.\ Rev.\ D {\bf 80}, 017301 (2009)
[arXiv:0808.0945 [hep-ph]].

\bibitem{Ananthanarayan:2012onz}
B.~Ananthanarayan, S.~K.~Garg, M.~Patra and S.~D.~Rindani,
Phys.\ Rev.\ D {\bf 85}, 034006 (2012)
[arXiv:1104.3645 [hep-ph]].

\bibitem{Ananthanarayan:2014cal}
B.~Ananthanarayan, J.~Lahiri, M.~Patra and S.~D.~Rindani,
JHEP {\bf 08}, 124 (2014)
[arXiv:1404.4845 [hep-ph]].

\bibitem{Rahaman:2016nzs}
R.~Rahaman and R.~K.~Singh,
Eur.\ Phys.\ J.\ C {\bf 76}, 539 (2016)
[arXiv:1604.06677 [hep-ph]].

\bibitem{Rahaman:2017qed}
R.~Rahaman and R.~K.~Singh,
Eur.\ Phys.\ J.\ C {\bf 77}, 521 (2017)
[arXiv:1703.06437 [hep-ph]].

\bibitem{Ellis:2020ekm}
J.~Ellis, S.~F.~Ge, H.~J.~He and R.~Q.~Xiao,
Chin.\ Phys.\ C {\bf 44}, 063106 (2020)
[arXiv:1902.06631 [hep-ph]].

\bibitem{Fu:2021jec}
Q.~Fu, J.~C.~Yang, C.~X.~Yue and Y.~C.~Guo,
Nucl.\ Phys.\ B {\bf 972}, 115543 (2021)
[arXiv:2102.03623 [hep-ph]].

\bibitem{Ellis:2021rop}
J.~Ellis, H.~J.~He and R.~Q.~Xiao,
Sci.\ China\ Phys.\ Mech.\ Astron. {\bf 64}, 221062 (2021)
[arXiv:2008.04298 [hep-ph]].

\bibitem{Yang:2022tgw}
J.~C.~Yang, Y.~C.~Guo and L.~H.~Cai,
Nucl.\ Phys.\ B {\bf 977}, 115735 (2022)
[arXiv:2111.10543 [hep-ph]].

\bibitem{Spor:2022pou}
S.~Spor, E.~Gurkanli and M.~K{\"o}ksal,
Nucl.\ Phys.\ B {\bf 979}, 115785 (2022)
[arXiv:2203.02352 [hep-ph]].

\bibitem{Jahedi:2022duc}
S.~Jahedi and J.~Lahiri,
JHEP \textbf{04} (2023), 085
doi:10.1007/JHEP04(2023)085
[arXiv:2212.05121 [hep-ph]].

\bibitem{Jahedi:2023myu}
S.~Jahedi,
JHEP \textbf{12} (2023), 031
doi:10.1007/JHEP12(2023)031
[arXiv:2305.11266 [hep-ph]].

\bibitem{Baur:1993fkx}
U.~Baur and E.~L.~Berger,
Phys.\ Rev.\ D {\bf 47}, 4889 (1993).

\bibitem{Senol:2018gvg}
A.~Senol, H.~Denizli, A.~Yilmaz, I.~T.~Cakir, K.~Y.~Oyulmaz, O.~Karadeniz and O.~Cakir,
Nucl.\ Phys.\ B {\bf 935}, 365-376 (2018)
[arXiv:1805.03475 [hep-ph]].

\bibitem{Rahaman:2019tnp}
R.~Rahaman and R.~K.~Singh,
Nucl.\ Phys.\ B {\bf 948}, 114754 (2019)
[arXiv:1810.11657 [hep-ph]].

\bibitem{Senol:2019ybv}
A.~Senol, H.~Denizli, A.~Yilmaz, I.~T.~Cakir and O.~Cakir,
Acta\ Phys.\ Pol.\ B {\bf 50}, 1597 (2019)
[arXiv:1906.04589 [hep-ph]].

\bibitem{Senol:2020hbh}
A.~Senol, H.~Denizli, A.~Yilmaz, I.~T.~Cakir and O.~Cakir,
Phys.\ Lett.\ B {\bf 802}, 135255 (2020)
[arXiv:1910.03843 [hep-ph]].

\bibitem{Yilmaz:2020ser}
A.~Yilmaz, A.~Senol, H.~Denizli, I.~T.~Cakir and O.~Cakir,
Eur.\ Phys.\ J.\ C {\bf 80}, 173 (2020)
[arXiv:1906.03911 [hep-ph]].

\bibitem{Yilmaz:2021dbm}
A.~Yilmaz,
Nucl.\ Phys.\ B {\bf 969}, 115471 (2021)
[arXiv:2102.01989 [hep-ph]].

\bibitem{Hernandez:2021wsz}
A.~I.~Hern{\'a}ndez-Ju{\'a}rez, A.~Moyotl and G.~Tavares-Velasco,
Eur.\ Phys.\ J.\ C {\bf 81}, 304 (2021)
[arXiv:2102.02197 [hep-ph]].

\bibitem{Biekotter:2021ysx}
A.~Biek{\"o}tter, P.~Gregg, F.~Krauss and M.~Sch{\"o}nherr,
Phys.\ Lett.\ B {\bf 817}, 136311 (2021)
[arXiv:2102.01115 [hep-ph]].

\bibitem{Lombardi:2022plb}
D.~Lombardi, M.~Wiesemann, G.~Zanderighi,
Phys.\ Lett.\ B {\bf 824}, 136846 (2022).

\bibitem{Hernandez-Juarez:2022kjx}
A.~I.~Hern\'andez-Ju\'arez and G.~Tavares-Velasco,
[arXiv:2203.16819 [hep-ph]].

\bibitem{Ellis:2022zdw}
J.~Ellis, H.~J.~He and R.~Q.~Xiao,
Phys. Rev. D \textbf{107} (2023) no.3, 035005
doi:10.1103/PhysRevD.107.035005
[arXiv:2206.11676 [hep-ph]].

\bibitem{Ellis:2023ucy}
J.~Ellis, H.~J.~He and R.~Q.~Xiao,
Phys. Rev. D \textbf{108} (2023) no.11, L111704
doi:10.1103/PhysRevD.108.L111704
[arXiv:2308.16887 [hep-ph]].

\bibitem{Chapon:2009hh}
E.~Chapon, C.~Royon and O.~Kepka,
Phys. Rev. D \textbf{81} (2010), 074003
doi:10.1103/PhysRevD.81.074003
[arXiv:0912.5161 [hep-ph]].

\bibitem{Kepka:2008yx}
O.~Kepka and C.~Royon,
Phys. Rev. D \textbf{78} (2008), 073005
doi:10.1103/PhysRevD.78.073005
[arXiv:0808.0322 [hep-ph]].

\bibitem{Geng:2019ebo}
C.~Geng {\it et al.} [ATLAS Collaboration],
PoS {\bf DIS2019}, 286 (2019).

\bibitem{Gounaris:2003lsd}
G.~J.~Gounaris, J.~Layssac and F.~M.~Renard,
Phys.\ Rev.\ D {\bf 67}, 013012 (2003)
[arXiv:hep-ph/0211327].

\bibitem{Belloni:2022due}
A.~Belloni, A.~Freitas, J.~Tian, J.~Alcaraz Maestre, A.~Apyan, B.~Azartash-Namin, P.~Azzurri, S.~Banerjee, J.~Beyer and S.~Bhattacharya, \textit{et al.}
[arXiv:2209.08078 [hep-ph]].

\bibitem{Acar:2016rde}
Y.~C.~Acar, A.~N.~Akay, S.~Beser, A.~C.~Canbay, H.~Karadeniz, U.~Kaya, B.~B.~Oner and S.~Sultansoy,
Nucl. Instrum. Meth. A \textbf{871} (2017), 47-53
doi:10.1016/j.nima.2017.07.041
[arXiv:1608.02190 [physics.acc-ph]].

\bibitem{Canbay:2017rbg}
A.~C.~Canbay, U.~Kaya, B.~Ketenoglu, B.~B.~Oner and S.~Sultansoy,
Adv. High Energy Phys. \textbf{2017} (2017), 4021493
doi:10.1155/2017/4021493
[arXiv:1704.03534 [physics.acc-ph]].

\bibitem{Ketenoglu:2022fzo}
B.~Keteno\u{g}lu, B.~Da\u{g}l\i{}, A.~\"Ozt\"urk and S.~Sultansoy,
Mod. Phys. Lett. A \textbf{37} (2022) no.37n38, 2230013
doi:10.1142/S0217732322300130

\bibitem{Delahaye:2019egb}
J.~P.~Delahaye, M.~Diemoz, K.~Long, B.~Mansouli{\'e}, N.~Pastrone, L.~Rivkin, D.~Schulte, A.~Skrinsky and A.~Wulzer,
“Muon Colliders,” (2019)
[arXiv:1901.06150 [physics.acc-ph]].

\bibitem{Long:2021wja}
K.~R.~Long, D.~Lucchesi, M.~A.~Palmer, N.~Pastrone, D.~Schulte and V.~Shiltsev,
Nat.\ Phys. {\bf 17}, 289-292 (2021).

\bibitem{CEPC}
 The CEPC Study Group, [arXiv:1811.10545 [hep-ex]].

\bibitem{Palmer:2014asd}
R.~B.~Palmer,
Rev.\ Accel.\ Sci.\ Tech. {\bf 7}, 137-159 (2014).

\bibitem{Antonelli:2016ezx}
M.~Antonelli, M.~Boscolo, R.~D.~Nardo and P.~Raimondi,
Nucl.\ Instrum.\ Meth.\ A {\bf 807}, 101-107 (2016)
[arXiv:1509.04454 [physics.acc-ph]].

\bibitem{Wang:2016rwe}
M-H.~Wang, Y.~Nosochkov, Y.~Cai and M.~Palmer,
JINST {\bf 11}, P09003 (2016).

\bibitem{Neuffer:2018zxp}
D.~Neuffer and V.~Shiltsev,
JINST {\bf 13}, T10003 (2018)
[arXiv:1811.10694 [physics.acc-ph]].

\bibitem{Boscolo:2019ytr}
M.~Boscolo, J-P.~Delahaye and M.~Palmer,
Rev.\ Accel.\ Sci.\ Tech. {\bf 10}, 189-214 (2019)
[arXiv:1808.01858 [physics.acc-ph]].

\bibitem{Bogomilov:2020twm}
B.~Bogomilov {\it et al.} [MICE Collaboration],
Nature {\bf 578}, 53-59 (2020)
[arXiv:1907.08562 [physics.acc-ph]].

\bibitem{Buttazzo:2018wmg}
D.~Buttazzo, D.~Redigolo, F.~Sala and A.~Tesi,
JHEP {\bf 11}, 144 (2018)
[arXiv:1807.04743 [hep-ph]].

\bibitem{Koksal:2019lja}
M.~K{\"o}ksal, A.~A.~Billur, A.~Guti{\'e}rrez-Rodr{\'{\i}}guez and M.~A.~Hern{\'a}ndez-Ru{\'{\i}}z,
Int.\ J.\ Mod.\ Phys.\ A {\bf 34}, 1950076 (2019)
[arXiv:1811.01188 [hep-ph]].

\bibitem{Costantini:2020tkp}
A.~Costantini, F.~D.~Lillo, F.~Maltoni, L.~Mantani, O.~Mattelaer, R.~Ruiz and X.~Zhao,
JHEP {\bf 09}, 80 (2020)
[arXiv:2005.10289 [hep-ph]].

\bibitem{Yin:2020gre}
W.~Yin and M.~Yamaguchi,
``Muon $g-2$ at multi-TeV muon collider,'' (2020)
[arXiv:2012.03928 [hep-ph]].

\bibitem{Ruhdorfer:2020tgx}
M.~Ruhdorfer, E.~Salvioni and A.~Weiler,
SciPost\ Phys. {\bf 8}, 027 (2020)
[arXiv:1910.04170 [hep-ph]].

\bibitem{Chiesa:2020yhn}
M.~Chiesa, F.~Maltoni, L.~Mantani, B.~Mele, F.~Piccinini and X.~Zhao,
JHEP {\bf 09}, 98 (2020)
[arXiv:2003.13628 [hep-ph]].

\bibitem{Bandyopadhyay:2021lja}
P.~Bandyopadhyay and A.~Costantini,
Phys.\ Rev.\ D {\bf 103}, 015025 (2021)
[arXiv:2010.02597 [hep-ph]].

\bibitem{Han:2021hrq}
T.~Han, S.~Li, S.~Su, W.~Su and Y.~Wu,
Phys.\ Rev.\ D {\bf 104}, 055029 (2021)
[arXiv:2102.08386 [hep-ph]].

\bibitem{Liu:2021gtr}
W.~Liu and K-P.~Xie,
JHEP {\bf 04}, 15 (2021)
[arXiv:2101.10469 [hep-ph]].

\bibitem{Han:2021twq}
T.~Han, Z.~Liu, L.-T.~Wang and X.~Wang,
Phys.\ Rev.\ D {\bf 103}, 075004 (2021)
[arXiv:2009.11287 [hep-ph]].

\bibitem{Capdevilla:2021xku}
R.~Capdevilla, F.~Meloni, R.~Simoniello and J.~Zurita,
JHEP {\bf 06}, 133 (2021)
[arXiv:2102.11292 [hep-ph]].

\bibitem{Bottaro:2021res}
S.~Bottaro, A.~Strumia and N.~Vignaroli,
JHEP {\bf 06}, 143 (2021)
[arXiv:2103.12766 [hep-ph]].

\bibitem{Capdevilla:2021ooc}
R.~Capdevilla, D.~Curtin, Y.~Kahn and G.~Krnjaic,
Phys.\ Rev.\ D {\bf 103}, 075028 (2021)
[arXiv:2006.16277 [hep-ph]].

\bibitem{Huang:2021edc}
G-Y.~Huang, F.~S.~Queiroz and W.~Rodejohann,
Phys.\ Rev.\ D {\bf 103}, 095005 (2021)
[arXiv:2101.04956 [hep-ph]].

\bibitem{Asadi:2021wsd}
P.~Asadi, R.~Capdevilla, C.~Cesarotti and S.~Homiller,
JHEP {\bf 10}, 182 (2021)
[arXiv:2104.05720 [hep-ph]].

\bibitem{Han:2021pas}
T.~Han, D.~Liu, I.~Low and X.~Wang,
Phys.\ Rev.\ D {\bf 103}, 013002 (2021)
[arXiv:2008.12204 [hep-ph]].

\bibitem{Franceschini:2021pol}
R.~Franceschini and M.~Greco,
Symmetry {\bf 13}, 851 (2021)
[arXiv:2104.05770 [hep-ph]].

\bibitem{Arakawa:2022mkr}
J.~Arakawa, A.~Rajaraman, T.~Sui and T.~M.~P.~Tait,
SciPost Phys. \textbf{16} (2024) no.3, 072
doi:10.21468/SciPostPhys.16.3.072
[arXiv:2208.14464 [hep-ph]].

\bibitem{Chiesa:2021tyr}
M.~Chiesa, B.~Mele and F.~Piccinini,
``Multi Higgs production via photon fusion at future multi-TeV muon colliders,'' (2021)
[arXiv:2109.10109 [hep-ph]].

\bibitem{Buttazzo:2021eka}
D.~Buttazzo and P.~Paradisi,
Phys.\ Rev.\ D {\bf 104}, 075021 (2021)
[arXiv:2012.02769 [hep-ph]].

\bibitem{Huang:2022vke}
G-Y.~Huang, S.~Jana, F.~S.~Queiroz and W.~Rodejohann,
Phys.\ Rev.\ D {\bf 105}, 015013 (2022)
[arXiv:2103.01617 [hep-ph]].

\bibitem{Spor:2022kyz}
S.~Spor and M.~K{\"o}ksal,
CJP {\bf 101}, 10 (2023)
[arXiv:2201.00787 [hep-ph]].

\bibitem{Yang:2022dbn}
J-C.~Yang, X-Y.~Han, Z-B.~Qin, T.~Li and Y-C.~Guo,
JHEP {\bf 2022}, 74 (2022)
[arXiv:2204.10034 [hep-ph]].

\bibitem{Forslund:2022unz}
M.~Forslund and P.~Meade,
JHEP {\bf 2022}, 185 (2022)
[arXiv:2203.09425 [hep-ph]].

\bibitem{Cepedello:2024ogz}
R.~Cepedello, F.~Esser, M.~Hirsch and V.~Sanz,
JHEP \textbf{07} (2024), 275
[erratum: JHEP \textbf{12} (2024), 084]
doi:10.1007/JHEP07(2024)275
[arXiv:2402.04306 [hep-ph]].

\bibitem{Ellis:2024omd}
J.~Ellis, H.~J.~He, R.~Q.~Xiao, S.~P.~Zeng and J.~Zheng,
[arXiv:2408.12508 [hep-ph]].

\bibitem{Acar:2017eli}
Y.~C.~Acar, U.~Kaya and B.~B.~Oner,
Chin. Phys. C \textbf{42} (2018) no.8, 083108
doi:10.1088/1674-1137/42/8/083108
[arXiv:1703.04030 [hep-ph]].

\bibitem{Aydin:2021iky}
G.~Aydin, Y.~O.~G\"unaydin, M.~T.~Tarakcioglu, M.~Sahin and S.~Sultansoy,
Acta Phys. Polon. B \textbf{53} (2022) no.11, 3
doi:10.5506/APhysPolB.53.11-A3
[arXiv:2105.09686 [hep-ph]].

\bibitem{Caliskan:2018vep}
A.~Caliskan,
APhysPolB {\bf 50} 8
[arXiv:1802.09874 [hep-ph]].

\bibitem{Degrande:2014ydn}
C.~Degrande,
JHEP {\bf 02}, 101 (2014)
[arXiv:1308.6323 [hep-ph]].

\bibitem{Grzadkowski:2010es}
B.~Grzadkowski, M.~Iskrzynski, M.~Misiak and J.~Rosiek,
JHEP \textbf{10} (2010), 085
doi:10.1007/JHEP10(2010)085
[arXiv:1008.4884 [hep-ph]].

\bibitem{Buchmuller:1985jz}
W.~Buchmuller and D.~Wyler,
Nucl. Phys. B \textbf{268} (1986), 621-653.

\bibitem{Gounaris:2000svs}
G.~J.~Gounaris, J.~Layssac and F.~M.~Renard,
Phys.\ Rev.\ D {\bf 61}, 073013 (2000).

\bibitem{Rahaman:2020fdf}
R.~Rahaman,
Indian Institute of Science Education and Researh, PhD thesis (2020).

\bibitem{Degrande:2013kka}
C.~Degrande,
JHEP \textbf{02} (2014), 101
doi:10.1007/JHEP02(2014)101.

\bibitem{Aaboud:2018ybz}
M.~Aaboud {\it et al.} [ATLAS Collaboration],
JHEP {\bf 12}, 010 (2018).

\bibitem{Alwall:2014cvc}
J.~Alwall, R.~Frederix, S.~Frixione, V.~Hirschi, F.~Maltoni, O.~Mattelaer, H.~S.~Shao, T.~Stelzer, P.~Torrielli and M.~Zaro,
JHEP {\bf 07}, 079 (2014).

\bibitem{Weizsacker} C. F. von Weizsacker, {\it Z. Phys.} {\bf 88}, 612 (1934).

\bibitem{Williams} E. J. Williams, {\it Kong. Dan. Vid. Sel. Mat. Fys. Med.} {\bf 13N4}, 1 (1935).

\bibitem{Budnev} V. M. Budnev, I. F. Ginzburg, G. V. Meledin and V. G. Serbo, {\it Phys. Rep.} {\bf 15}, 181 (1975).

\bibitem{Chen2} M. S. Chen, T. P. Cheng, I. J. Muzinich and H. Terazawa, {\it Phys. Rev.} {\bf D7}, 3485 (1973).

\bibitem{Eboli4} O. J. P. Eboli, M. C. Gonzalez-Garcia, and S. M. Lietti, {\it Phys. Rev.} {\bf D69}, 095005 (2004).

\bibitem{Ellison-ARNPS1998} J. Ellison and J. Wudka, {\it Annu. Rev. Nucl. Part. Sci.} {\bf 48}, 33 (1998).

\bibitem{khoriauli} G. Khoriauli, {\it Nuovo Cimento B} {\bf 123}, 1327-1330 (2008).

\bibitem{CMS:2020gtj}
A.~M.~Sirunyan \textit{et al.} [CMS],
Eur. Phys. J. C \textbf{81} (2021) no.3, 200

\bibitem{emre}
A.~Senol, S.~Spor, E.~Gurkanli, V.~Cetinkaya, H.~Denizli  and M.~Koksal,
Eur.\ Phys.\ J.\ P {\bf 137}, 1354 (2022).

\end{thebibliography}
\end{document}